\documentclass[11pt,a4paper]{article}
\usepackage[T1]{fontenc}
\usepackage{graphicx}
\begin{document}

\title{ Undissociated screw dislocations in silicon: Calculations of core structure and energy }

\author{ L. Pizzagalli\footnote{Email: Laurent.Pizzagalli@univ-poitiers.fr, Phone: +33 549 496833, Fax: +33 549 496692},
P. Beauchamp, and J. Rabier
\\ \\ Laboratoire de M\'etallurgie Physique\\
Unit\'e Mixte de Recherche 6630 du CNRS, Universit\'e de Poitiers, \\
B.P. 30179 \\
86962 Futuroscope Chasseneuil Cedex, France}

\date{}

\maketitle
\begin{abstract}

The stability of the perfect screw dislocation in silicon has been investigated using both classical potentials 
and first principles calculations. Although a recent study stated that the stable screw was located 
both in the 'shuffle' and 'glide' sets of \{111\} planes (Koizumi \textit{et al}, 2000, Phil. Mag. A, 80, 609), 
it is shown that this result depends on the classical potential used, and 
that the most stable configuration belongs to the 'shuffle' set only, in the centre of one $(\bar{1}01)$
hexagon. We also investigated the stability of an sp$^2$ hybridization in the core of the dislocation, obtained for one
metastable configuration in the 'glide' set. The core structures are characterized in several ways, with a description of the three
dimensional structure, differential displacement maps, and derivatives of the dis-registry. 


\end{abstract}


\section{Introduction} \label{intro}

Dislocations in silicon have been the subject of many investigations, both experimental, computational and theoretical, not
only because they can appear in microelectronic devices, but also because of their own properties, closely related to the covalent 
nature of bonding in this material. In ordinary conditions, silicon is 
brittle below about 600$^\circ$C (Hirsch, Samuels and Roberts 1989, George and Michot 1993). Above this temperature, TEM observations
show dissociated dislocations (Ray and Cockayne 1971) which thus lie between the narrowly  spaced \{111\} planes \textit{i.e.} belonging 
to the glide set. On the basis of computed generalized stacking-fault energy and entropy calculations, on \{111\} narrowly 
spaced (glide set) and widely spaced (shuffle set) respectively (Kaxiras and Duesbery 1993), the idea was put forward of the possibility of 
a transition : at low temperature, perfect dislocations are easier to nucleate and move in the shuffle set, while the 
activation barrier for glide becomes lower for dissociated dislocations in the glide set at high temperature (Duesbery and Joos 1996). 

Thus, experiments have been undertaken in which, in order to achieve plastic deformation at temperatures as low as room temperature,
the silicon sample is either submitted to a high confining pressure, of the order of 5~GPa (Rabier \textit{et al} 2001) or subject to a 
surface scratch test (Rabier \textit{et al} 2000). Under these conditions, the microstructure has been found formed of
non-dissociated dislocations, supposed to belong to the shuffle set. Favoured dislocation orientations appear to be 
screw, 60$^\circ$, 30$^\circ$ and also 41$^\circ$. Similarly, deformation experiments in III-V compounds such as GaAs, 
InP, InSb, performed down to 77K by applying a high confining pressure (Suzuki \textit{et al} 1998, 1999a, b), 
indicate that the low temperature plastic deformation is governed by kink pair formation on undissociated screw dislocations 
moving in the shuffle set planes.

Theoretical investigations of the core structures of dislocations in silicon are then clearly required to bring additional 
insights. However, despite a large number of existing atomistic computations, most of these were devoted to partial 
dislocations of the glide set, with a particular attention to core reconstructions of the 30$^\circ$ and 90$^\circ$ 
partials and to mobility properties (cf. for instance the review by Bulatov \textit{et al} 2001). Less information 
is available about perfect dislocations:
In his pioneer examination of dislocation cores in diamond cubic structures Hornstra (1958) 
quite naturally placed the screw dislocation line furthest away from any atom row, that is at 
the center of the hexagon formed by six neighbouring <110> dense atom rows; the dislocation then 
belongs to the shuffle set, more precisely it is located at the intersection of two \{111\} 
shuffle planes at 71.53$^\circ$. By somewhat artificially rebonding atoms, Hornstra also proposed 
another structure for the screw core, the interesting idea being that the core is spread 
over two adjacent hexagons sharing a common small edge.  Arias and Joannopoulos (1994) performed 
DFT calculations of the shuffle screw in silicon: they found the first configuration proposed by 
Hornstra to be stable with respect to spontaneous dissociation and they calculated energy parameters 
for the core. Finally, in a recent study, Koizumi, Kamimura and Suzuki (2000) have investigated the 
core configuration and the mobility of the (a/2)<110> screw using the Stillinger-Weber (1985) potential 
for silicon. They found two stable configurations: the configuration at the centre of 
the hexagon, denoted A, has a higher energy and the lower energy configuration (denoted B) can be 
regarded as belonging to both a \{111\} shuffle plane and the \{111\} glide plane at 71.53$^\circ$. The 
authors discussed in detail the very special part that configuration B might play in cross-slip mechanism and 
in the transition of dislocation glide from the shuffle set to the glide set. It remains to be confirmed 
if these results are general and not specific to  the Stillinger-Weber potential. Regarding the  glide set, 
previous works essentially focused on partial dislocations, and, to our knowledge, there are no 
available studies of the perfect screw configurations. 

Thus, it is of interest to investigate the core properties of perfect dislocations in the shuffle and glide 
sets, and particularly the existence and relative stability of all proposed configurations of the screw 
orientation. This paper reports on such calculations, using both (i) the Density Functional Theory (DFT) 
formalism and (ii) several semi-empirical potentials. After a description of the 
methods and computational details, we present the different energetic and structural parameters associated with each configuration. 
These quantities are then discussed in relation with the previous results, dislocation mobilities and validity of the 
classical potentials.

\section{Computational methods}

The atomistic calculations have first been carried out  using semi-empirical potentials. We have employed 
(i) the SW potential of Stillinger-Weber (1985), used as reference and also for comparison with previous work from 
Koizumi \textit{et al} (2000), (ii) the Tersoff potential (1988) which is able to give a 
better representation of a number of defects than Stillinger-Weber, and (iii) EDIP (Justo \textit{et al} 1998) 
constructed so as to benefit from the successes of earlier potentials and incorporating data 
obtained from DFT calculations, such as the gamma-surfaces. It has to be noted that for dislocation calculations, EDIP 
is the only semi-empirical potential able to account for reconstructions of both 30$^\circ$ and 90$^\circ$ 
partials in the glide set. The main advantage of empirical potentials is their low computational cost, which 
allows for a  fast calculation of several configurations. Potentials 
suffer from limitations, implicitly related to their functional form or limited fitting database, and 
calculated energies may prove to be relatively inaccurate, especially for configurations involving 
highly distorted or broken bonds, as encountered in dislocation cores. However, by using three different types 
of potentials, we expect to overcome this issue and obtain reliable results.

First principles calculations have also been carried out, 
in order to discriminate between configurations, and to calculate more precise defect energies. In addition, 
comparisons with empirical potential allowed us to check their reliability for dislocation core investigations. We performed DFT
calculations (Hohenberg and Kohn 1964, Kohn and Sham 1965) in the local density approximation at zero temperature with the ABINIT
code (ABINIT, 2002). The ionic interactions were represented by norm-conserving pseudopotentials 
(Trouiller and Martins 1991). We used a plane-wave basis with 
an energy cutoff of 10~Ry and two special $k$-point along the dislocation line (Monkhorst and Pack, 1976). 
Tests with Generalized Gradient Approximations (GGA), a higher energy cutoff or a finer $k$-point sampling have also been 
conducted for selected configurations. We found an error on the defect energy lower than 0.5\% using 5~special k-points and 
a cutoff of 14~Ry, and about 5\% when using GGA. 

At first, semi-empirical potential calculations were made, for a fast exploration of several system sizes, thus 
determining the size effects of the simulation slab on the results. Suitable simulation box sizes  were then selected, 
small enough for DFT calculations to remain tractable and large enough for the computed energies to 
be meaningful.

\section{Simulation model}

Ideally, we would consider an isolated straight screw dislocation in an infinite bulk. Along the dislocation line, provided that 
there is no reconstruction, we have a periodic situation with 
a period equal to the Burgers vector $\frac{1}{2}[110]a_0$. The use of periodic boundary 
conditions is then the obvious suitable choice along the dislocation line. 
In the plane perpendicular to the dislocation line, a long-range strain field will be generated by the dislocation, and should be 
taken into account in the calculation. 
Two different methods could be employed. In the first one, no periodic boundary conditions perpendicular to the dislocation 
line are applied, and only one dislocation is located in the centre of the simulation box (figure~1A). 
The atomic positions at the boundaries are then initialized to values 
calculated with elasticity theory using a numerical code adapted from ANCALC (Stroh, 1958,1962) or with a more 
precise model (Lehto and \"{O}berg 1998). 
The computational box has to be large enough to 
prevent a fictitious interaction between fixed boundaries and the dislocation core. 
In addition, atoms located at the edges of the system are not in a bulk-like environment, 
and defect energies can not be directly extracted from total energy calculations. Here, we used such 
an approach for semi-empirical potential calculations only, because the simulation box could be enlarged at will, due to the 
low computational cost, and also because defined individual atomic energy allows an easy determination of defect energies. Typical 
computational cells involved about 10000 atoms (dimensions 133~\AA$\times$132~\AA$\times$11.5~\AA\ for a ($40\times84\times3$) cell). 
The anisotropic elastic energy per unit length of the dislocation is given by the well known formula (Hirth and Lothe 1982) 

$$
E = \frac{K\mathbf{b}^2}{4\pi}\ln\left(\frac{R}{r_0}\right)
$$

The second method involves periodic boundary conditions along the directions 
perpendicular to the dislocation line. In that case, annoying difficulties arise owing to discontinuities at the 
boundaries, in particular for a single dislocation. 
Spurious shear strain associated with these discontinuities could then have a significant influence on the 
dislocation core structure and energetics. These difficulties can be smoothed by considering dipolar 
(figure~1B) or quadrupolar (figure~1C)  
arrangements of dislocations in the cell. Whether a dipole or a quadrupole should be favoured depends on the 
character of the dislocation: a dipole is best suited for an edge 
dislocation whereas a quadrupole minimizes the residual strain associated with a quadrupole of screw 
dislocations (Lehto and \"{O}berg 1998). With the second choice, four 
dislocations should be included in the cell, with separation distances large enough to prevent a spurious interaction, which 
would lead to large cell sizes. However, as suggested by Bigger \textit{et al} (1992), the system could be divided by a 
factor of two by relaxing the orthogonality constraints on the periodic 
cell (figure~1D). In this work, the quadrupolar arrangements of dislocations (figure~1C and 1D) have been considered for both 
semi-empirical and first principles calculations. The semi-empirical calculations were useful for investigating easily several  
cell sizes and estimating the non-elastic core-core energy contributions possibly present for very small cells. 
We considered computational cells ranging from ($40\times84\times3$) to ($6\times12\times3$). For ab initio calculations, the 
largest ($12\times12\times1$) cell encompasses 144 atoms, with 2 dislocations.

For a quadrupolar distribution and four dislocations in an orthogonal cell, the anisotropic elastic energy per unit length
is obtained by summation of the interactions between dislocation pairs, calculated using a code adapted from ANCALC (Stroh, 1958,1962).
The total energy can be split up into an interaction energy inside the cell ($E^{\mbox{intra}}$) and half the interaction energy
between the quadrupole and all its periodically repeated images ($E^{\mbox{inter}}$). 
A reference (zero) of the elastic energy is required for determining $E^{\mbox{intra}}$, and is chosen 
as the elastic energy of a dislocations quadrupole whose distance along the edge is $d_0$. In that case, 
it can be shown that the reference distance $d_0$ is equal to 
the core radius $r_0$ obtained for a single dislocation. In fact, if the quadrupole is extremely large 
such that the four dislocations can be considered as isolated, the elastic energy amounts to four times the 
self energy of a single dislocation.
The determination of $E^{\mbox{inter}}$ 
should require an infinite summation, which has to be dealt with care. A quick convergence may be obtained by summing the 
interaction energy between quadrupoles. It has to be noted that 
this is not the case for a dipolar arrangement, and special handling of the summation is required (Cai \textit{et al} 2001). 
The derivation of the elastic energy for a quadrupolar distribution in a non-orthogonal cell (two dislocations per 
cell instead of four, see figure~1) is straightforward.

\section{Results}

The table~1 shows the elastic constants calculated with the semi-empirical potentials and first principles. 
These constants are used for generating the initial configurations from anisotropic elasticity theory, and 
extracting core energetics 
from relaxed systems. 

In figure~2, we show a $(\bar{1}01)$ section of the cubic diamond structure, with three possible locations of the 
dislocation line. A corresponds to the original position at the center of one hexagon (Hornstra 1958), for a screw dislocation 
belonging to two shuffle planes. B was recently proposed by Koizumi \textit{et al} (2000), at the middle of one 
long hexagon bond. It is interesting to point out that 
in this case, the dislocation is located at the crossing of both a shuffle and a glide $\{111\}$ plane. 
Finally, another high symmetry location on the structure is point C, at the middle of a short hexagon bond, with the 
screw dislocation belonging to two glide planes. Other locations have been investigated, either inside the hexagon
or at the exact position of one Si atom, but in all cases, the system relaxed to one of the three selected configurations.

The different energetic values resulting from all our calculations are reported in table~2. The energy differences show 
that with ab initio and all potentials but SW, A is the most stable configuration. We were able to reproduce the results of  
Koizumi \textit{et al} (2001), B appearing more stable than A by using the SW potential.
B is obtained as the second choice with ab initio and EDIP potential, while 
it seems highly unfavourable 
with the Tersoff potential. Another important point concerns the stability of the B configuration. 
Although the relaxation with semi-empirical potentials was straightforward, the B geometry has been found extremely difficult to
retain within a first principles 
calculation, even using initial configurations relaxed with potentials as a starting point. Some or all dislocations of the 
quadrupole generally evolved to an A configuration, or annihilated themselves. Only in one case were we able to 
relax the structure. Finally, for all calculations, C is never the most stable configuration, or is even unstable with the SW
potential. 

Table~2 also reports the core radii obtained by matching the elastic energy with the calculated defect energy. For the 
empirical potential calculations and one unique dislocation in the computational cell, the core radius $r_0$ is determined 
by considering the defect energy contained in the cylinder centered on the dislocation line, with radius $R$ and height 
the Burgers vector $\mathbf{b}$ (see formula above). The factor $K$ determined from the calculated elastic constants is used. 
We considered that the core radius $r_0$ is already well converged for $R=60$~\AA. For example, the core radius 
changes by less than 
0.01~\AA\ for $R$ ranging between 40 and 60~\AA, for a configuration A and the SW potential. 
For first principles calculations and a quadrupolar distribution of dislocations, the core radius is determined numerically by inverting
the defect elastic energy. 
We found that the core radius determination is already very precise for a ($12\times12$) cell, with an uncertainty about 
0.01~\AA. For the smaller ($6\times6$) cell, a 0.1~\AA\ deviation from converged values was obtained. Most of the 
values are close to about 1~\AA, the commonly used 1/4 of the Burgers vector. Only core radii for B and C with Tersoff and A 
with EDIP are slightly distant. In the table~2 are also reported core energy values, often used in the literature, which are 
obtained with a core radius equal to the Burgers vector.

In figure~3, differential displacement maps of the relaxed A, B and C configurations of the 
screw dislocations, obtained from systems with one unique dislocation and relaxed with an empirical potential.
For the configuration A, the distortion is uniformly distributed on the hexagon ring encircling the 
dislocation line. It is also clear from the picture that the displacements are identical along the two 'shuffle' planes (see 
figure~2). The configuration B is characterized by a maximal distortion on the two atoms on both sides of the 'shuffle' plane. 
Most of the constraints are located on the hexagon rings sharing these atoms. In the case of C, the maximal distortion is also
located on two atoms, but on both sides of a 'glide' plane. Average deformations are observed for all four hexagon 
rings around these two atoms, and equivalent displacements along the two glide planes passing through the C screw core.  
It is noteworthy that for each geometry, almost identical pictures have been obtained regardless of the potential 
considered. One exception is the C configuration with the SW potential, which relaxes to the C' configuration represented  
in figure~3. The initial differential displacement located on the two atoms next to the 
dislocation location vanished, and large distortions appeared involving more distant atoms (see figure~3). We found this
configuration to be 
unstable with both EDIP and Tersoff, relaxing to A. In figure~4, the differential displacement map for a
quadrupolar distribution of A configuration, relaxed with first principles, is represented. Even if the dislocation cores are close and 
interact together, it is noteworthy that the general pattern obtained for a unique dislocation remains easily recognizable in 
that case. 

To characterize the spatial extension of the dislocation core, we have determined the width at half maximum (WHM) of the derivative of 
the dis-registry introduced by the dislocation. The dis-registry is obtained by computing the difference of displacements along 
$(\bar{1}01)$ for atoms on either sides of the glide  $(\bar{1}1\bar{1})$ plane (Figure~5). The calculated points
are fitted on the simple shape $\arctan\left(\frac{x}{\Delta}\right)$ of a dislocation (Hirth and Lothe, 1982). The determination of the WHM 
is then straightforward. In table~3, we report all values, as well as the WHM calculated in the same way for 
configurations built from anisotropic elasticity theory. In contrast to the energy ordering, it appears that the 
WHM's do not depend on the kind of classical potential, for a given configuration. For C, the ab initio WHM is also equal to the 
classical one. It is also clear that the B core is wider than the A core. 

\section{Discussion}

The A configuration has been considered as the most plausible structure for the undissociated screw dislocation in silicon. But a recent study 
by Koizumi \textit{et al} (2000) concluded that the B configuration is slightly more stable than A. Our calculations indicate that A is
definitely the most stable geometry for the screw, using ab initio and two classical potentials. However, with the SW potential, 
we were able to reproduce the result of Koizumi \textit{et al}, which proves that the apparent stability of B over A is an artefact of  
the potential. With first principles calculations involving various cell sizes, the B configuration was found to be stable 
only in one case. It seems that the slightest deformation could lead to the relaxation from B to A. We conclude that the B 
configuration is weakly metastable.
Additional insights are obtained from the analysis of the core. The figure~6 shows three dimensional structures of three 
configurations. A is characterized by the absence of atomic rearrangements in the core, the main distortions being 
located on all bonds forming the hexagon ring encircling the dislocation core. On the contrary, for B, the main deformations are 
located on the two atoms close to the core. They are bonded together and at the same height along $(\bar{1}01)$ in the bulk. After 
introduction of the dislocation, the atoms have now a height difference of half the Burgers vector along 
$(\bar{1}01)$, and they are separated by about 2.8~\AA. Weak bonds with such an interatomic distance are possible for silicon. 
However, each of these atoms has already a coordination~3 and would need to accommodate 2~extra bonds (see figure~6), which  
is unlikely to occur. From the analysis of the electronic density, it appears that in the B core, 2 rows of dangling bonds follow 
the dislocation line. On the one hand, 
these rows may explain the very low stability of this configuration with first principles methods. On the other hand, the large range 
of energy values illustrates the difficulty of describing dangling bonds with the classical potentials. We investigated a possible 
reconstruction of the B configuration along the dislocation line, for reducing the number of dangling bonds. Because the atoms 
involved are located on either side of the 'shuffle' plane, new bonds are formed only at the expense of breaking other bonds. 
Interestingly, this by-hand reconstruction relaxed to the C configuration. The WHM, and so the core 
extension, of B is larger than A, which may also explain why it is less stable. 

We also investigated the possibility of a screw dislocation in the glide set with the C configuration. 
With all classical potentials and 
first principles method, it is found that A is more stable than C, with a large energy difference. It is then unlikely that the C core 
could be formed in bulk silicon. However, interesting features are associated with this structure. The examination of the geometry 
in the core revealed that the two atoms on either side of the dislocation line (black balls in figure~6) have a coordination~3.
Before the introduction of the dislocation, these two atoms, were bonded together, with a height difference of half the Burgers vector
along $(\bar{1}01)$. After, they are still bonded together but located at the same height, one bond per atom being broken. The 
ab initio interatomic distance between these two atoms is 2.16~\AA, whereas distances with neighbour atoms are about 2.29~\AA. 
The three bonds for each atom are almost co-planar, and the angle between them range from 117$^\circ$ to 123$^\circ$. All these 
quantities point at an sp$^2$ hybridization of these two atoms, with a double bond between them, which is confirmed by the 
analysis of the electronic density. This possibility has already been proposed by Hornstra (1958), on the basis of geometrical
arguments only. An sp$^2$ 
hybridization is not favoured in silicon, which explained the large defect energy for the C configuration. However, it would be
interesting to investigate the competition between screw configurations in diamond for example, where sp$^2$ is favoured over 
sp$^3$. It is worth noting that a sp$^2$ character is also present in recently proposed metastable structures for the  
30$^\circ$ and 90$^\circ$ partial dislocations in diamond (Ewels \textit{et al} 2001, Blumenau \textit{et al} 2002).
One interesting aspect of the C configuration is the tightness of the core. It is difficult to compare directly A and C since they 
are not located in the same family of {111} planes. Nevertheless, insights could be gained by the comparison with the anisotropic
elastic solution (Table~3). It is clear that for all potentials, and especially for ab initio, the relaxed A core is wider than 
the initial elastic configuration. The effect is even stronger for the B configuration. However, in the case of C, 
the core is contracted by the atomic relaxation. The narrowness of C could be partly attributed to the formation of the double 
bond in the core. 

To our knowledge, the A configuration has already been investigated with first principles techniques in two previous studies. 
Arias and Joannopoulos (1994) found a core energy $E_c=0.56\pm0.21$~eV.\AA$^{-1}$ whereas Miyata and Fujiwara (2001) obtained $E_c$ 
equal to 0.95~eV.\AA$^{-1}$. These results may be compared to our ab initio core energy of 0.52~eV.\AA$^{-1}$ (table~2), 
very close to the value of the former study. However, this agreement seems fortuitous, the authors using isotropic theory and 
a fitted $K\equiv\mu=0.29$~eV.\AA$^{-3}$. Miyata and Fujiwara (2001) followed a similar approach, but with a fitted 
$K\equiv\mu=0.48$~eV.\AA$^{-3}$, and obtained a larger core energy. Instead, here, $K$ factors calculated within 
anisotropic theory and with the ab initio calculated elastic constants, very close to the experimental values (see table~2), 
are employed. The disagreement between the different works may be explained by the poor k-point sampling in the newer study, 
as well as the use of isotropic theory with a fitted $K$ factor. 

We discussed our results in relation to the screw mobility. A being the most stable geometry, possible paths from one 
minimum to another include saddle configurations B or C (see Figure~2). Koizumi \textit{et al} (2000) found a Peierls stress  
of about 2~GPa for the non-dissociated screw dislocation, considering the path A$\rightarrow$B$\rightarrow$A. However, this 
relatively low value may be explained by the use of the SW potential, and its failure to yield the correct stability of A and 
B configurations. In fact, although the Peierls stress cannot be simply 
determined from static calculations, several insights may be obtained from the analysis of the 
energy differences between configurations. With SW, the energy difference between A and B is only 0.14~eV per Burgers vector. 
With first principles, we determined a larger energy difference of 0.32~eV per Burgers vector. It is then reasonable to assume that the 
Peierls stress for the path A$\rightarrow$B$\rightarrow$A will be much higher than 2~GPa. This is confirmed by recent ab initio 
calculations from Miyata and Fujiwara (2001), where the Peierls stress ranged from 22 to 30~GPa. Another possible path for 
dislocation cross slip would be A$\rightarrow$C$\rightarrow$A. However, the large calculated energy difference indicates a very 
large stress, and this possibility may be ruled out solely on the basis of energy considerations. 

Finally, we compared the merits of each classical potentials we have used in this study. On the basis of our three investigated configurations, it appears 
that EDIP is better suited that SW or the Tersoff potential for this study, the stability and energy differences being close to the ab 
initio results. This is not completely surprising since this potential has been designed specifically
to study defects (Justo 1998). The worst is maybe the SW potential, which yields the B core as the most stable configuration. 
It is worth noting that although the stabilities of the different 
core configurations very much depend on the kind of potential, we obtained similar atomic structures in almost all cases. 
Consequently, for relaxing configurations prior to ab initio calculations, one could use any classical potential. But for 
investigating stability, it is necessary to consider several kind of potentials.

\section{Conclusion}

Using anisotropic elasticity theory, several semi-empirical classical potentials, and first principles calculations, we have investigated the 
properties of the undissociated screw dislocation in silicon. Considering previous studies and the geometry of the silicon 
atomic structure, three possible structures have been selected and compared. We have shown that the configuration A, where 
the dislocation core is located in the centre of one hexagon, in the shuffle set, is clearly more stable than the two 
others. In a former study by Koizumi and Kamimura (2000), another configuration, with the dislocation located in the 
centre of one long hexagon edge, was favoured. From our calculations, it appears that 
this result is explained by the use of the Stillinger Weber
potential, this configuration being less stable with other classical potentials or ab initio methods. 
We also investigated a third solution, with the dislocation in the glide set. Despite its high defect 
energy, this configuration presents the interesting feature of an sp$^2$ hybridization of the atoms forming the core. 
Obviously, such a structure is worth to be studied in a material favouring sp$^2$, like diamond.  
We also characterized the spatial extension of the cores of each structure by determining the derivative of the dis-registry.  
A possible continuation of this work includes the determination of the mobility of the undissociated screw dislocation in silicon. 
The study of other dislocation orientations, such as those recently observed at low temperature (Rabier \textit{et al}, 2000, 2001), would be 
another working direction.

\section*{References}

{\ } 

ABINIT The ABINIT code is a common project of the Universite
       Catholique de Louvain, Corning Incorporated, and other contributors 
       (URL http://www.pcpm.ucl.ac.be/abinit).
   
Arias, T.A. and Joannopoulos, J.D., 1994, Phys. Rev. Let., 73, 680.

Bigger, J.R.K., McInnes, D.A., Sutton, A.P., Payne, M.C., Stich, I., King-Smith, R.D., Bird, D.M., and Clarke, L.J., 
1992, Phys. Rev. Lett. 69, 2224.

Blumenau, A. T., Heggie, M. I., Fall, C. J., Jones, R., and Frauenheim, T., 2002, Phys. Rev. B, 65, 205205.

Bulatov, V.V., Justo, J.F., Cai, W., Yip, S., Argon, A.S., Lenosky, T., de Koning, M., and Diaz de la Rubia, T., 2001, Phil. Mag. A,
81, 1257.

Cai, W., Bulatov, V.V., Chang, J., Li, J., and Yip, S., 2001, Phys. Rev. Lett., 86, 5727.

Duesbery, M.S. and Joos, B., 1996, Phil. Mag. A, 74, 253.

Ewels, C. P., Wilson, N. T., Heggie, M. I., Jones, R., and Briddon, P. R., 2001, J. Phys.: Condens. Matter, 13, 8965. 

George, A., and Michot, G., 1993, Mater.  Sci.  Engng., A164, 118.

Hirsch, P.B., Roberts, S.G., and Samuels, S., 1989, Proc. R. Soc.  London A 421, 25.

Hirth, J. P., and Lothe, J., 1982, Theory of Dislocation (New York: Wiley). 

Hohenberg, P. and Kohn, W., 1964, Phys. Rev., 136, B 864.

Hornstra, J., 1958, J. Phys. Chem. Solids, 5, 129.

Justo, J.F., Bazant, M.Z., Kaxiras, E., Bulatov, V.V. and Yip, S. 1998, Phys. Rev. B, 58, 2539.

Kaxiras, E. and Duesbery, M.S., 1993, Phys. Rev. Lett., 70, 3752.

Kohn, W. and Sham, L.J., 1965, Phys. Rev., 140, A 1133.

Koizumi, H., Kamimura, Y., and Suzuki, T., 2000, Phil. Mag. A, 80, 609.

Lehto, N., and \"{O}berg, S., 1998, Phys. Rev. Lett., 80, 5568.

Miyata, M., and Fujiwara, T., 2001, Phys. Rev. B, 63, 045206.

Monkhorst, H.J., and Pack, J.D., 1976, Phys. Rev. B, 13, 5188.

Rabier, J., Cordier, P., Tondellier, T., Demenet, J.L., and Garem, H., 2000, J. Phys.: Condens. Matter., 49, 10059

Rabier, J., Cordier, P., Demenet, J.L., and Garem, H., 2001, Mat. Sci. Eng A, 309-310, 74

Ray ,I.L.F., and Cockayne, D.J.H., 1971, Proc. R. Soc. London, A325, 543.

Simmons, G., and Wang H., 1971, Single Crystal Elastic Constants and Calculated Aggregate 
Properties: A Handbook (Cambridge, MA: MIT)

Stillinger, F.H., and Weber, T.A., 1985, Phys. Rev. B, 31, 5262. 

Stroh, A.N., 1958, Phil. Mag., 3, 625; 1962, J. math. Phys., 41, 77

Suzuki, T., Nishisako, T., Taru, T., and Yasutomi, T., 1998, Phil. Mag. Let., 77, 173. 

Suzuki, T. Yasutomi, T., Tokuoka, T., and Yonenaga, I., 1999a, Phys. stat. sol. (a), 171, 47; 1999b, Phil. Mag. A, 79, 2637.

Tersoff, J., 1988, Phys. Rev. B, 38, 9902.

Trouiller, N., and Martins, J.L., 1991, Phys. Rev. B, 43, 1993.

\newpage 

\section*{Table caption}

\begin{table}[h]
\caption{Experimental (Simmons and Wang 1971) and calculated elastic constants (in Mbar) for the SW Stillinger-Weber 
(Stillinger and Weber 1985), Tersoff (Tersoff 1989) and EDIP (Justo \textit{et al} 1998) potentials as well as our ab initio results.
The SW parameters have been rescaled in order to fit the experimental cohesive energy of 4.63 eV. For a screw dislocation in 
a cubic diamond structure, $K=\left[C_{44}(C_{11}-C_{12})/2\right]^{-\frac{1}{2}}$}
\end{table}

\begin{table}[h]
\caption{Calculated energetic parameters for A, B and C screw dislocations. $\Delta E$ is the energy difference, in eV per 
Burgers vector, with an uncertainty of 0.01~eV/Bv. $r_0$ is the core radius ($\pm 0.03$~\AA). $E_c$ is the core energy, obtained 
with a fixed core radius equal to the Burgers vector ($\pm 0.02$~eV.\AA$^{-1}$). Note that C is not stable 
with the SW potential, it relaxed to a configuration with an energy 0.02 eV/Bv higher than A.}
\end{table}

\begin{table}[h]
\caption{WHM (in \AA) obtained from elasticity (with experimental $C_{ij}$), classical potentials, and ab initio. Two values are undetermined, 
due to the instability of C with SW, and the difficulty to obtain B with first principles in a large cell.}
\end{table}

\clearpage

\ \vspace{4cm}

\begin{table}[h]
\begin{center}

\begin{tabular}{@{\extracolsep{10pt}}*{6}{c}}
\hline
\hline
  & exp. & SW & Tersoff & EDIP & ab initio \\
\hline
$B$ & 0.99 & 1.083 & 0.978 & 0.99 & 0.99\\
$C_{11}$ & 1.67 & 1.617 & 1.425 & 1.75 & 1.64 \\
$C_{12}$ & 0.65 & 0.816 & 0.754 & 0.62 & 0.66 \\
$C_{44}$ & 0.81 & 0.603 & 0.687 & 0.71 & 0.78 \\
$C_{44}^0$ &   & 1.172 & 1.188 & 1.12 & 1.09 \\
$K$ & 0.64 & 0.49 & 0.48 & 0.63 & 0.62 \\
\hline
\hline
\end{tabular}
\end{center}\end{table}

\vfill\centerline{\bf Table 1}

\clearpage

\ \vspace{4cm}

\begin{table}[h]
\begin{center}

\begin{tabular}{@{\extracolsep{10pt}}*{9}{c}}
\hline
\hline
  & \multicolumn{2}{c}{$\Delta E$ (eV/Bv)} & \multicolumn{3}{c}{Radius $r_0$ (\AA)} & 
  \multicolumn{3}{c}{Energy $E_c$ (eV.\AA$^{-1}$)} \\
  & $E_B-E_A$ & $E_C-E_A$ & A & B & C & A & B & C \\
\hline
 SW & -0.14 &   & 0.82 & 0.91  &   & 0.55 & 0.51 &   \\
 Tersoff & 1.08 & 0.54 & 0.78 & 0.35  & 0.52 & 0.55 & 0.84 & 0.70 \\
 EDIP & 0.23 & 0.74 & 1.49 & 1.31 & 0.98 & 0.37 & 0.42 & 0.54 \\
ab initio & 0.32 & 0.86 & 1.22 & 1.03  & 0.74 & 0.52 &  0.60 & 0.74 \\
\hline
\hline
\end{tabular}
\end{center}\end{table}

\vfill\centerline{\bf Table 2}

\clearpage

\ \vspace{4cm}

\begin{table}[h]
\begin{center}
\begin{tabular}{@{\extracolsep{10pt}}*{6}{c}}
\hline
\hline
  & Elastic & SW & Tersoff & EDIP & ab initio \\
\hline
A & 2.7 & 3.1 & 3.2 & 3.2 & 3.6\\
B & 2.6 & 3.9 & 4.0 & 4.0 &   \\
C & 1.2 &     & 0.9 & 0.9 & 0.9 \\
\hline
\hline
\end{tabular}
\end{center}\end{table}

\vfill\centerline{\bf Table 3}

\clearpage

\section*{Figure captions}

\noindent Figure~1: Models for straight dislocation simulations. In A, there is no periodic boundary conditions and only one 
dislocation in the computational cell (dashed red line). B, C and D show periodic boundary systems 
with dipolar (B) or quadrupolar (C,D) distributions of dislocations.  
\vspace{1cm}

\noindent Figure~2: Ball and stick representation of the $(\bar{1}01)$ plane of the cubic diamond structure. The three circles A, B and C mark the 
positions of the dislocation line. Dashed (dotted) lines show the 'shuffle' ('glide') $\{111\}$ planes.
\vspace{1cm}

\noindent Figure~3: Differential displacement maps of the screw dislocation in the configurations A, B, C, and C'
(obtained with SW from C). The arrows are proportional to the out-of-plane $[\bar{1}01]$ shifts between neighbour atoms   
introduced by the dislocation. The cross marks the position of the dislocation line 
and the dashed line the cut plane.
\vspace{1cm}

\noindent Figure~4: Differential displacement maps of the screw dislocation in the configuration A, in the quadrupolar distribution 
shown in the figure~1. The arrows are proportional to the out-of-plane $[\bar{1}01]$ shifts between neighbour atoms   
introduced by the dislocation.
\vspace{1cm}

\noindent Figure~5: Variation of the dis-registry in the (111) plane along the [121] direction for the three configurations. The 
solid lines are fits with the expression 
$f(x)=b\left[\frac{1}{\pi}\arctan\left(\frac{x}{\Delta}\right)-\frac{1}{2}\right]$. The insert graph 
shows the derivative of the dis-registry for A, and the definition of the WHM.
\vspace{1cm}

\noindent Figure~6: Ball-and-stick representation of the cubic diamond bulk (top left) and of the three screw core
configurations. A six atom ring (see figure~2) is represented by dark grey sticks, in order to show the deformation 
due to the dislocation and the Burgers vector. Dangling bonds (for B) and sp$^2$ atoms (for C) are represented by 
black sticks and balls. The position of the dislocation line is shown by the dashed lines.

\clearpage

\ \vspace{3cm}

\begin{figure}[h]
\includegraphics[width=14cm]{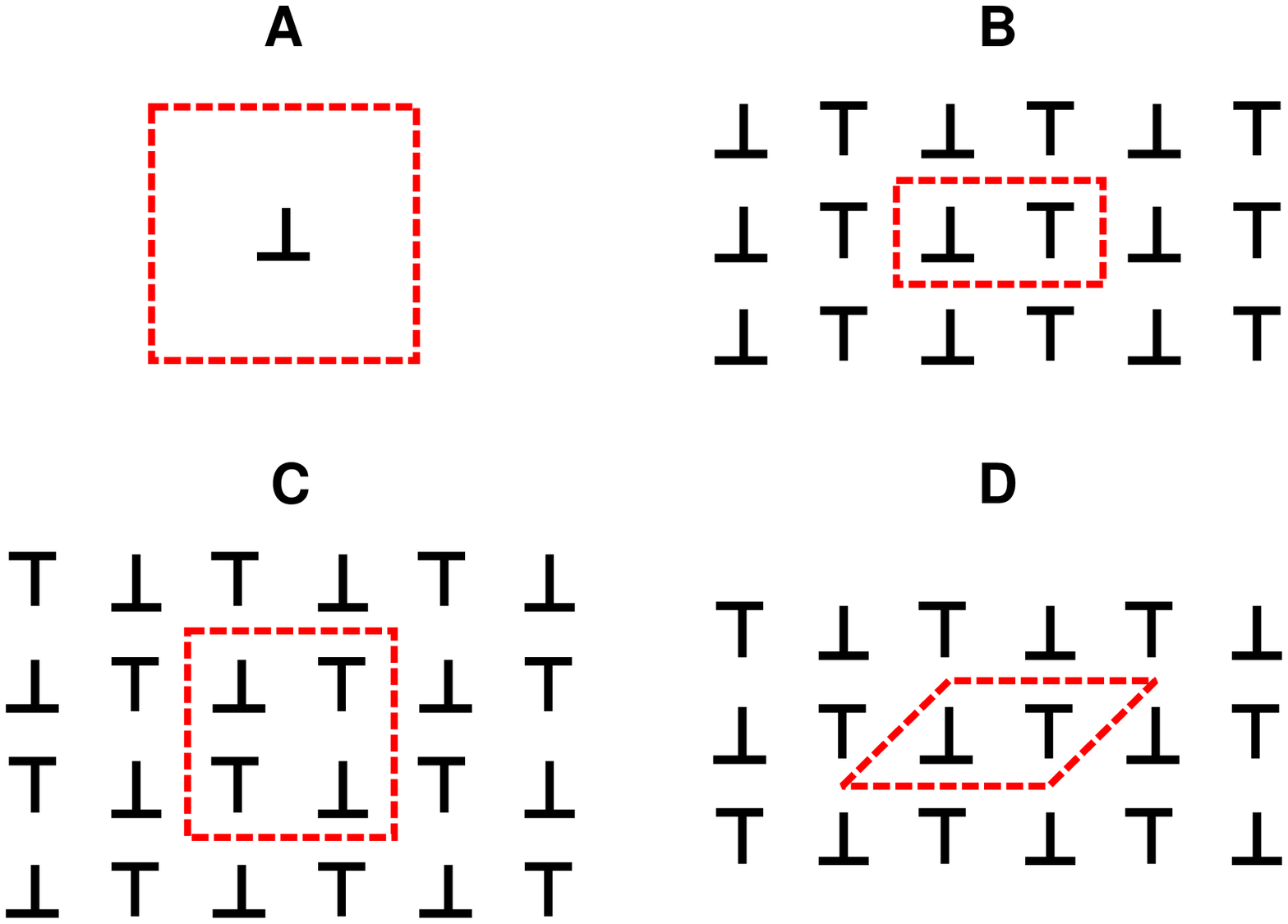}
\end{figure}

\vfill\centerline{\bf Figure 1}

\clearpage

\ \vspace{3cm}
\begin{figure}[h]
\includegraphics[angle=-90,width=15cm]{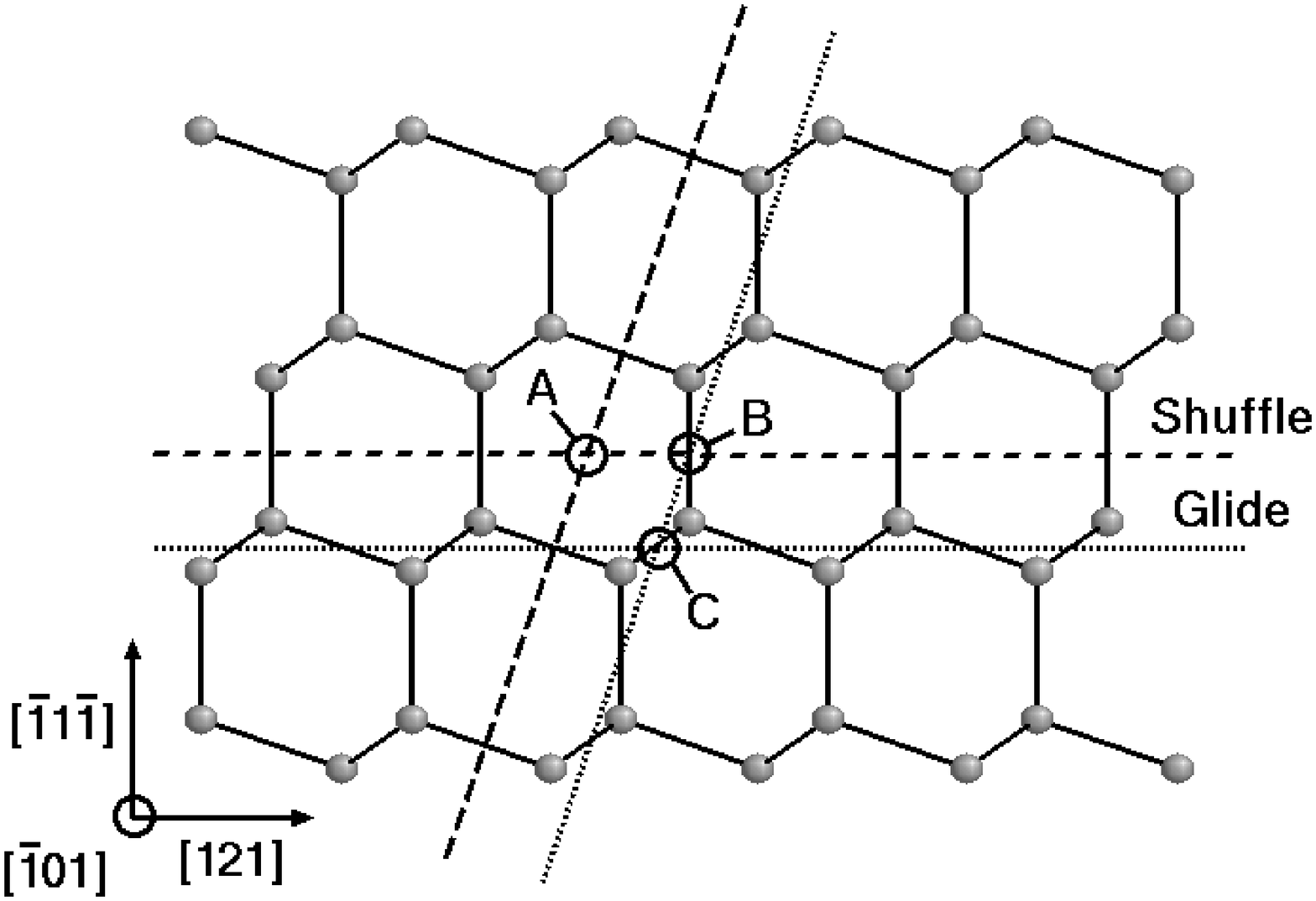}
\end{figure}

\vfill\centerline{\bf Figure 2}

\clearpage

\ \vspace{2cm}
\begin{figure}[h]
\includegraphics[width=15.5cm]{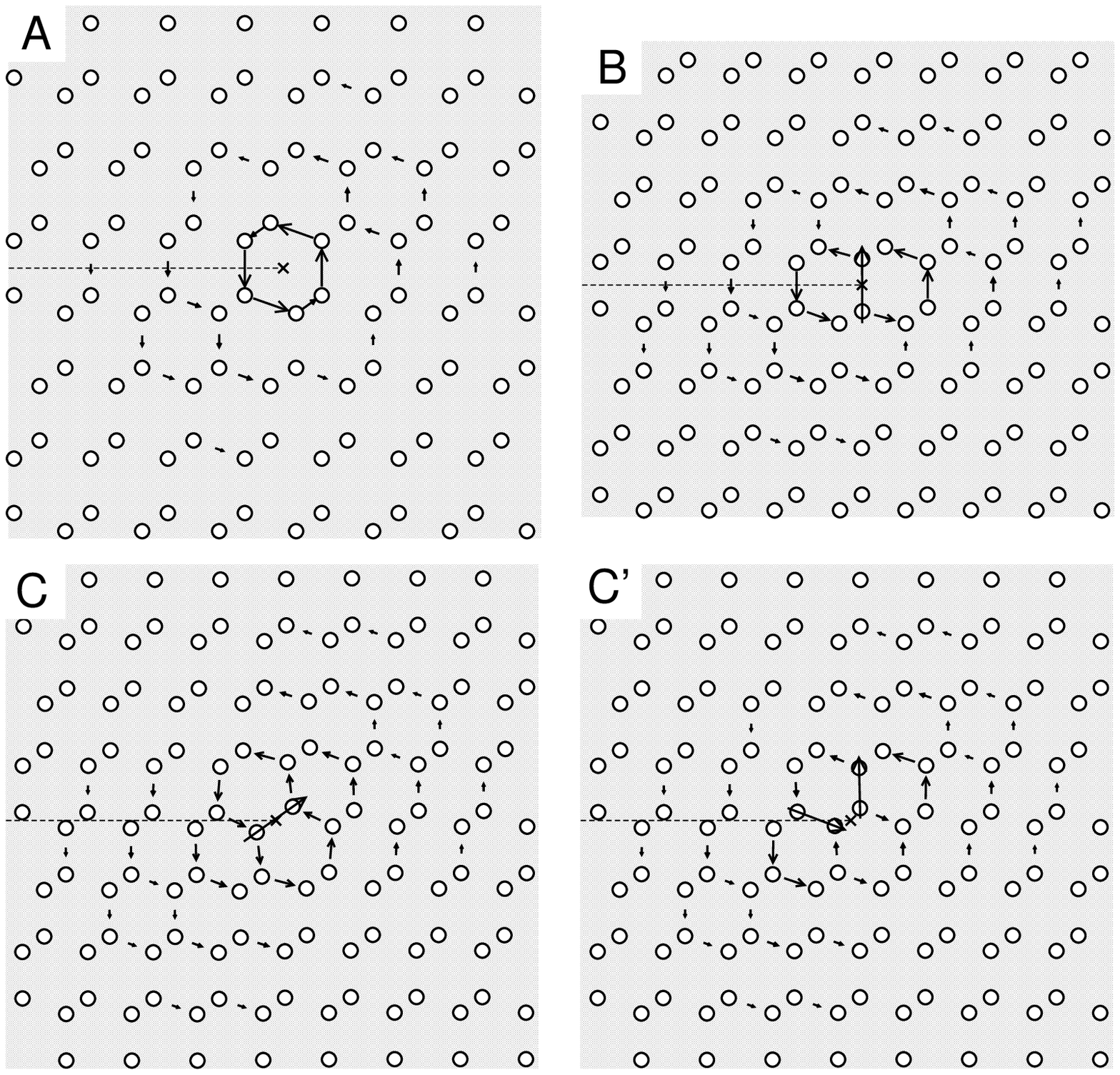}
\end{figure}

\vfill\centerline{\bf Figure 3}

\clearpage

\ \vspace{2cm}
\begin{figure}[h]
\includegraphics[width=15cm]{fig4.eps}
\end{figure}

\vfill\centerline{\bf Figure 4}

\clearpage

\ \vspace{2cm}
\begin{figure}[h]
\includegraphics[width=15.5cm]{fig5.eps}
\end{figure}

\vfill\centerline{\bf Figure 5}

\clearpage

\ \vspace{2cm}
\begin{figure}[h]
\includegraphics[angle=-90,width=15cm]{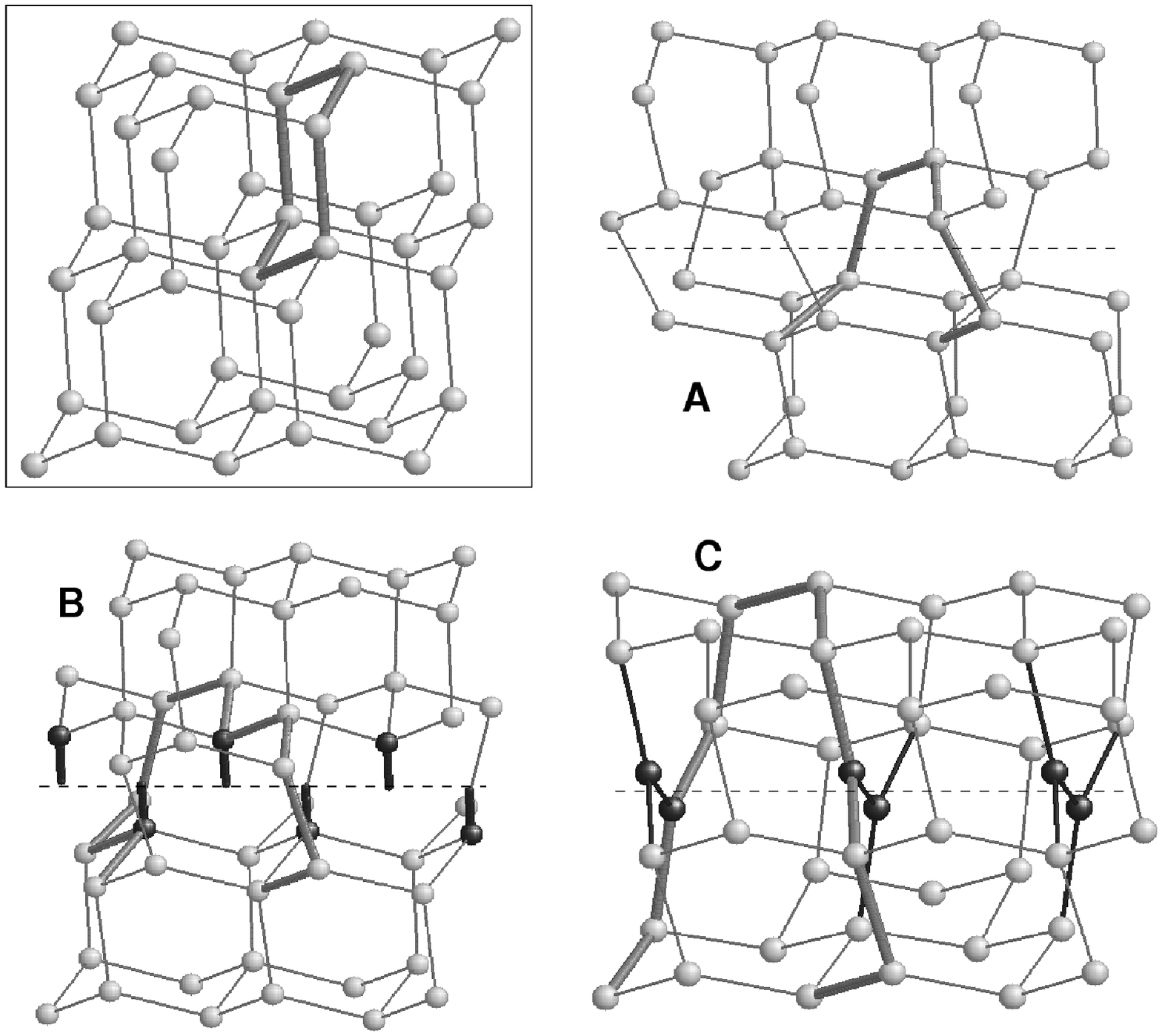}
\end{figure}

\vfill\centerline{\bf Figure 6}

\end{document}